\documentclass[aps,pra,showpacs,showkeys,twocolumn]{revtex4-1}
\usepackage{lineno,hyperref}
\modulolinenumbers[5]

\usepackage{color}
\usepackage{amsmath, amsthm, amssymb}
\usepackage{graphicx}
\usepackage{subfigure}
\usepackage{mathrsfs}

\usepackage[T1]{fontenc}
\usepackage[utf8]{inputenc}
\usepackage{lmodern}

\def\Tr{\,{\rm Tr}}

\def\i{\,{\rm i}}

\newtheorem{theorem}{Theorem}
\newtheorem{proposition}{Proposition}
\newtheorem{conjecture}{Conjecture}

\begin{document}


\title{Monoparametric family of metrics derived from classical Jensen-Shannon divergence}



\author{Trist\'an M. Os\'an}
\email{tosan@famaf.unc.edu.ar; tristan.osan@unc.edu.ar}
\affiliation{Instituto de F\'{\i}sica Enrique Gaviola (IFEG), Consejo Nacional de Investigaciones Cient\'{\i}ficas y T\'ecnicas de la Rep\'ublica Argentina (CONICET), Av. Medina Allende s/n, Ciudad Universitaria, X5000HUA, C\'ordoba, Argentina}
\affiliation{Facultad de Matem\'atica, Astronom\'{\i}a, F\'{\i}sica y Computaci\'on, Universidad Nacional de C\'ordoba, Av. Medina Allende s/n, Ciudad Universitaria, X5000HUA, C\'ordoba, Argentina}

\author{Diego G. Bussandri}
\affiliation{Facultad de Matem\'atica, Astronom\'{\i}a, F\'{\i}sica y Computaci\'on, Universidad Nacional de C\'ordoba, Av. Medina Allende s/n, Ciudad Universitaria, X5000HUA, C\'ordoba, Argentina}
\affiliation{Consejo Nacional de Investigaciones Cient\'{\i}ficas y T\'ecnicas de la Rep\'ublica Argentina (CONICET), Av. Rivadavia 1917, C1033AAJ, CABA, Argentina}

\author{Pedro W. Lamberti}
\affiliation{Facultad de Matem\'atica, Astronom\'{\i}a, F\'{\i}sica y Computaci\'on, Universidad Nacional de C\'ordoba, Av. Medina Allende s/n, Ciudad Universitaria, X5000HUA, C\'ordoba, Argentina}
\affiliation{Consejo Nacional de Investigaciones Cient\'{\i}ficas y T\'ecnicas de la Rep\'ublica Argentina (CONICET), Av. Rivadavia 1917, C1033AAJ, CABA, Argentina}

\begin{abstract}
Jensen-Shannon divergence is a well known multi-purpose measure of dissimilarity between probability distributions.
It has been proven that the square root of this quantity is a true metric in the sense that, in addition to the basic properties of a distance, it also satisfies the triangle inequality. In this work we extend this last result to prove that in fact it is possible to derive a monoparametric family of metrics from the classical Jensen-Shannon divergence. Motivated by our results, an application into the field of symbolic sequences segmentation is explored. Additionally, we analyze the possibility to extend this result into the quantum realm.
\keywords{Jensen-Shannon divergence, metrics, information theory, quantum distances}
\end{abstract}

%
%
%

\maketitle


\section{Introduction}

Measures of dissimilarity between probability distributions constitute an important topic of research in Probability Theory, Statistics and Information Geometry. Among some fields of application of this kind of measures we can mention, evaluation of risks in statistical decision problems, signal detection, data compression, coding, pattern classification, cluster analysis, etc. Furthermore, many problems of statistical physics can be established in terms of a measure of distance or distinguishability between two probability distributions.\\
In the realms of Statistics and Information Theory, an extensively applied measure of dissimilarity between probability distributions,  is the Jensen-Shannon divergence (JSD) \cite{Rao1987,Lin1991}. This measure turns out to be a symmetrized, smoothed, well-behaved and bounded version of the Kullback-Leibler divergence \cite{Kullback1951,Kullback1968}. JSD has been successfully applied in a wide variety of research fields, such as, analysis and characterization of symbolic sequences and segmentation of digital images. Particularly, it has been exhaustively used in the study of segmentation of DNA sequences. Remarkably, in statistical physics JSD has been used as a measure of the length of the time's arrow \cite{Feng2008} and also in the definition of a measure of complexity \cite{Rosso2007}. In addition, the generalization of JSD within the framework of the non-extensive Tsallis statistics \cite{Tsallis2009} has been studied in \cite{Lamberti2003,Majtey2004}.\\
In this work we show that it is possible to derive a monoparametric family of metrics from classical Jensen-Shannon divergence. A key aspect of our approach for the demonstration of this assertion is to consider the JSD as a particular case of a Csisz\'ar divergence \cite{Ali1966, Csiszar1963,Csiszar1967}.\\
In information geometry there exists a natural Riemannian structure associated with a local metric known as Fisher's metric \cite{Amari1993,Amari2016}. \v{C}encov showed that Fisher's metric is the only Riemannian metric on the probability distributions space $\mathscr{P}$ for which certain natural statistical embeddings are isometries \cite{Cencov1982}. Besides this formal fact, Fisher's metric is directly related with the practical parameters' estimation problem via the Cramer--Rao bound \cite{Amari1993,Amari2016}. Furthermore, the existence of Fisher's metric allows the space $\mathscr{P}$ to possess the character of metric space. Indeed, by evaluating the length of a geodesic associated with Fisher's metric, joining two points on the probability distributions space, we can provide a way of measuring the distance between two arbitrary points belonging to the space $\mathscr{P}$. It should be emphasized that we make a distinction between the local metric (which measures how separated are two near points from each other) and a measure of the distance between two arbitrary points. The distance defined through this procedure verifies the properties of a metric (cf. Sec. \ref{intrometrics} where we summarize these properties). It is worth to mention that the inverse procedure, i.e., to derive a Riemannian metric from a metric, is not always possible. 
Additionally, it has been shown that having a metric defined on $\mathscr{P}$ (and over any arbitrary space) is of crucial importance to establish convergence criteria in iterative processes \cite{Bryant1985}.\\
This paper is organized as follows. In Sec. \ref{theory}, we introduce the basic theoretical background related to our work. Next, in Sec. \ref{mainresult} we prove the main result of this work, i.e., that it is possible to construct a monoparametric family of metrics from the classical expression of JSD. Then, in Sec. \ref{applications} we briefly explore the possibilities of applying our results in two different contexts. On one hand, in Sec. \ref{segmentation} we explore the segmentation of symbolic sequences. On the other hand, in Sec. \ref{Qrealm} we study the extension of the monoparametric family of metrics into the quantum realm. Finally, in Sec. \ref{concl} we summarize our results.

\section{Theoretical framework\label{theory}}

\subsection{Divergences, distances and metrics\label{intrometrics}}

From a mathematically rigorous viewpoint, a \textit{metric} (or sometimes, a \textit{true} metric) $d$ on a set $\chi$ is a function $d:\chi \times \chi \rightarrow \mathbb{R}_{\geq 0}$ such that for
any $x,y,z \in \chi$ the following properties are satisfied
\begin{enumerate}
\item \label{M1} {\em Non-negativity}: $d(x,y) \geq 0$
\item \label{M2} {\em Identity of indiscernibles}: $d(x,y) = 0$ if and only if $x=y$
\item \label{M3} {\em Symmetry}: $d(x,y) = d(y,x)$ 
\item \label{M4} {\em Triangle inequality}: $d(x,y) \leq d(x,z) + d(z,y)$
\end{enumerate}

In the context of classical information, usually $\chi$ represents the set of probability distributions and $x$ or $y$ represent an entire probability distribution such as $P = \{p_1, p_2,\cdots, p_n\}$ ($p_i \geq 0\; \forall i$, $\sum_{i=1}^n p_i = 1$). Often, if a distance measure $d$ only satisfies the property \ref{M1}, is called a \textit{divergence}. If, additionally, $d$ satisfies the properties \ref{M2} and \ref{M3} then $d$ is called a \textit{distance} \cite{Bryant1985,Hayashi2015,Deza2016}. It is worth to mention that throughout literature the term distance is used many times as equivalent to metric. Due to this use can be misleading in some contexts, for the sake of clarity, throughout this work we will use the terms divergence, distance or metric, according to the specific meaning needed.

\subsection{Csisz\'ar's divergences\label{sec:fdivdef}}

Csisz\'ar's divergences, also known as $f-$divergences, constitute an important class of measures of distinguishability between probability distributions \cite{Ali1966, Csiszar1963}. Let $\mathcal{F}$ be the set of convex functions  $f : \mathbb{R}_+ \mapsto \bar{\mathbb{R}}$ which are finite on $\mathbb{R}_0$ and continuous on $\mathbb{R}_+$, where $\bar{\mathbb{R}} = (-\infty,\infty]$, $\mathbb{R}_+ = [0,\infty)$ and $\mathbb{R}_0 = (0,\infty)$. The Csisz\'ar's $f-$divergence between the probability distributions $P = \{p_1, p_2,\cdots, p_n\}$ and $Q = \{q_1, q_2,\cdots, q_n\}$ is defined as \cite{Ali1966, Csiszar1963,Csiszar1967}

\begin{equation}
D_f(P,Q) = \sum_{i=1}^n q_i \, f\!\left(\frac{p_i}{q_i}\right)\label{fdivdef}
\end{equation}

\noindent From its definition, it can be seen that $D_f(P,Q)$ is a useful functional form which encompasses most of the commonly used divergence measures between probability distributions, such as, Kullback-Leibler divergence, Variational Distance, Hellinger distance, $\chi^2$-divergence, Jensen-Shannon divergence, among others \cite{Lin1991,Kullback1951,Kullback1968,Deza2016,Bengtsson2006}.

\subsubsection{Basic properties of Csisz\'ar's divergences\label{ciszarbasics}}

In what follows we will summarize some basic properties of Csisz\'ar's divergences directly related with the main result of this work. Further analysis of the properties of $f-$divergences can be found in references \cite{Vajda1972, Csiszar1974, Liese1987, Kafka1991,Oster2003}.

Let $f^* \in\mathcal{F}$, the *--conjugate (convex) function of $f$, be defined as:

\begin{equation}
f^*(u) = u\, f\!\left(\frac{1}{u}\right)\:\:\: \mathrm{for}\:\:\: u\in \mathbb{R}_0.
\end{equation}

If $f(1) = 0$, $f$ is strictly convex at $1$, and $f^*(u) = f(u)$, then $D_f(P,Q)$ satisfies the following basic properties \cite{Kafka1991,Oster2003,Oster1996}:

\begin{enumerate}
\item \label{P1} {\em Non-negativity and Identity of indiscernibles}: $D_f(P,Q) \geq 0$ with $D_f(P,Q) = 0$ $\iff P=Q$
\item \label{P2} {\em Symmetry}: $D_f(P,Q) = D_f(Q,P)$
\item \label{P3} {\em Uniqueness}: $D_{f_1}(P,Q) = D_f(P,Q)$, $\iff \exists c \in \mathbb{R}\,/\, f_1(u) = f(u) + c(u-1)$
\item \label{P4} {\em Range of values}: $f(1) \leq D_f(P,Q) \leq f(0) + f^*(0)$
\end{enumerate}

Bearing in mind the properties of a metric described in Sec. \ref{intrometrics} it can be seen that properties \ref{P1} and \ref{P2} are essential in order to seek for a metric based on a Csisz\'ar divergence. The following theorem addresses precisely the additional conditions that a Csisz\'ar divergence needs to satisfy in order for $[D_f(P,Q)]^\alpha$ to be a metric (cf. Sec. \ref{intrometrics}).

\begin{theorem}\label{Ostertheorem}
Let $D_f(P,Q)$ denote a Csisz\'ar divergence (cf. Sec. \ref{sec:fdivdef}  and Sec. \ref{ciszarbasics}). If $f(1) = 0$ and $f$ is strictly convex at $1$, $f^*(u) = f(u)$, and there exists $\alpha \in \mathbb{R}_0$ such that the function

\begin{equation}
h_\alpha(u)=\frac{\left(1-u^\alpha\right)^{1/\alpha}}{f(u)} \label{eq:metriccond}
\end{equation}

\noindent is nonincreasing for $u\in [0,1)$, then $d_\alpha(P,Q)\doteq [D_f(P,Q)]^\alpha$, in addition to properties \ref{P1} and \ref{P2}, also satisfies the triangle inequality $d_\alpha(P,Q) \leq d_\alpha(P,R) + d_\alpha(R,Q)$. Therefore, $d_\alpha(P,Q)$ is a metric \cite{Kafka1991,Oster2003,Oster1996}. $\blacksquare$
\end{theorem}

\subsubsection{KL divergence}

The Kullback-Leibler divergence (KLD) between two probability distributions $P = \{p_1,p_2,\cdots, p_n\}$ and $Q = \{q_1,q_2,\cdots, q_n\}$ is defined as \cite{Kullback1951,Kullback1968}:

\begin{equation}
D_{KL}(P,Q)=\sum^n_{i=1} p_i \log_2 \left(\frac{p_i}{q_i}\right)
\end{equation}

It is straightforward to show that $D_{KL}(P,Q)$ is a Csisz\'ar divergence (cf. eq. \eqref{fdivdef}) taking the function $f(u)$ as:

\begin{equation}
f_{KL}(u)=u\log_2(u)
\end{equation}

\noindent for $u\in \{\mathbb{R}_{+} \cup \{0\}\}$. 

The KLD constitutes a paradigmatic case of a widely used measure of dissimilarity between probability distributions which is not a metric. For example, $D_{KL}(P,Q)$ does not fulfill the basic property of symmetry \eqref{P2} essential for a metric. In addition, it should be noted that this divergence also possesses another undesired features. For example, if for any $j$ there exists a $q_j = 0$ for which $p_j \neq 0$, then $D_{KL}(P,Q)$ is undefined. Therefore, the probability distribution $P$ must be absolutely continuous with respect to the probability distribution $Q$ in order for $D_{KL}(P,Q)$ to be well defined \cite{Kullback1968}. However, it is worth to mention that this latter undesired feature of KLD can be avoided if the convention $0\, (-\infty) = 0$ is adopted.

\subsubsection{Classical Jensen-Shannon divergence}

The Jensen-Shannon divergence (JSD) between two probability distributions is defined as follows \cite{Rao1987,Lin1991}

\begin{equation}
D_{JS}(P,Q)= \frac{1}{2}\left[ D_{KL}\left(P,\frac{P+Q}{2}\right) + D_{KL}\left(Q,\frac{P+Q}{2}\right)  \right]\label{JSDdef} 
\end{equation}

After some algebra, a more explicit version of JSD can be written in terms of Shannon entropy $H = - \sum_i p_i \log_2 p_i$ as follows

\begin{equation}
D_{JS}(P,Q)= H\left(\frac{P+Q}{2}\right) - \frac{1}{2} H(P) - \frac{1}{2} H(Q)\label{JSDHVN}
\end{equation}

In last equation,  the dissimilarity between both probability distributions $P$ and $Q$ is evaluated  
assuming that both distributions have the same weight (1/2). If, instead, we consider arbitrary weights
$\pi_1\geq 0$ and $\pi_2 \geq 0 $, such that $\pi_1\,+\pi_2 = 1$, then equation \eqref{JSDHVN} can be generalized as follows \cite{Rao1987,Lin1991}:

\begin{equation}
D^{(\pi_1,\pi_2)}_{JS}(P,Q)= H\left(\pi_1\,P+\pi_2\,Q\right) - \pi_1 H(P) - \pi_2 H(Q)\label{JSDHVNgen}
\end{equation}

It is easy to prove that the expression \eqref{JSDHVN} can be written in the form of a Csisz\'ar divergence (cf. eq. \eqref{fdivdef}) with the function $f(u)$ defined as:

\begin{equation}
f_{JS}(u) = \frac{1}{2} \Big[ (1+u) + u \log_2(u) - (1+u) \log_2 (1+u) \Big]
\end{equation}
\noindent for $u\in \mathbb{R}_{+}$. 

From Eq. \eqref{JSDdef} it can be seen that $D_{JS}(P,Q)$ is a symmetric version of $D_{KL}(P,Q)$. Originated in the field of Information Theory \cite{Rao1987,Lin1991}, JSD is always well defined and bounded. Unlike KLD, as JSD satisfies the properties \ref{M1}, \ref{M2} and \ref{M3} (cf. Sec. \ref{intrometrics}), it is in fact a distance. Additionally, JSD has several interesting interpretations. For example, in statistical inference theory it gives both the lower and upper bounds to Bayes' probability error, whereas in the framework of information theory JSD can be related to mutual information \cite{Grosse2002}

\section{Monoparametric family of metrics from classical Jensen-Shannon divergence\label{mainresult}}

It is well known that the square root of the classical JSD, i.e., $[D_{JS}(P,Q)]^{1/2}$ (cf. Eq. \eqref{JSDHVN}) is a (true) metric, i.e., it satisfies the properties \ref{M1}, \ref{M2}, \ref{M3} and \ref{M4} (cf. Sec. \ref{intrometrics}) \cite{Oster1996,Endres2003}. In what follows we will extend this result to show that $d_\alpha(P,Q) \doteq [D_{JS}(P,Q)]^{\alpha}$ for $\alpha \in (0,1/2]$ is a (true) metric, i.e., we will prove that is possible to construct a monoparametric family of metrics from the classical expression of JSD. A key aspect of our approach for the proof is to consider the JSD as a particular example of a Csisz\'ar divergence. This allows us to use the theorem \ref{Ostertheorem} (cf. Sec. \ref{ciszarbasics}) to derive the conditions on the values of the parameter $\alpha$ for the function $d_\alpha(P,Q)$ to be a (true) metric.

\begin{figure}[h!]
\begin{center}
\includegraphics[clip,width=0.45\textwidth]{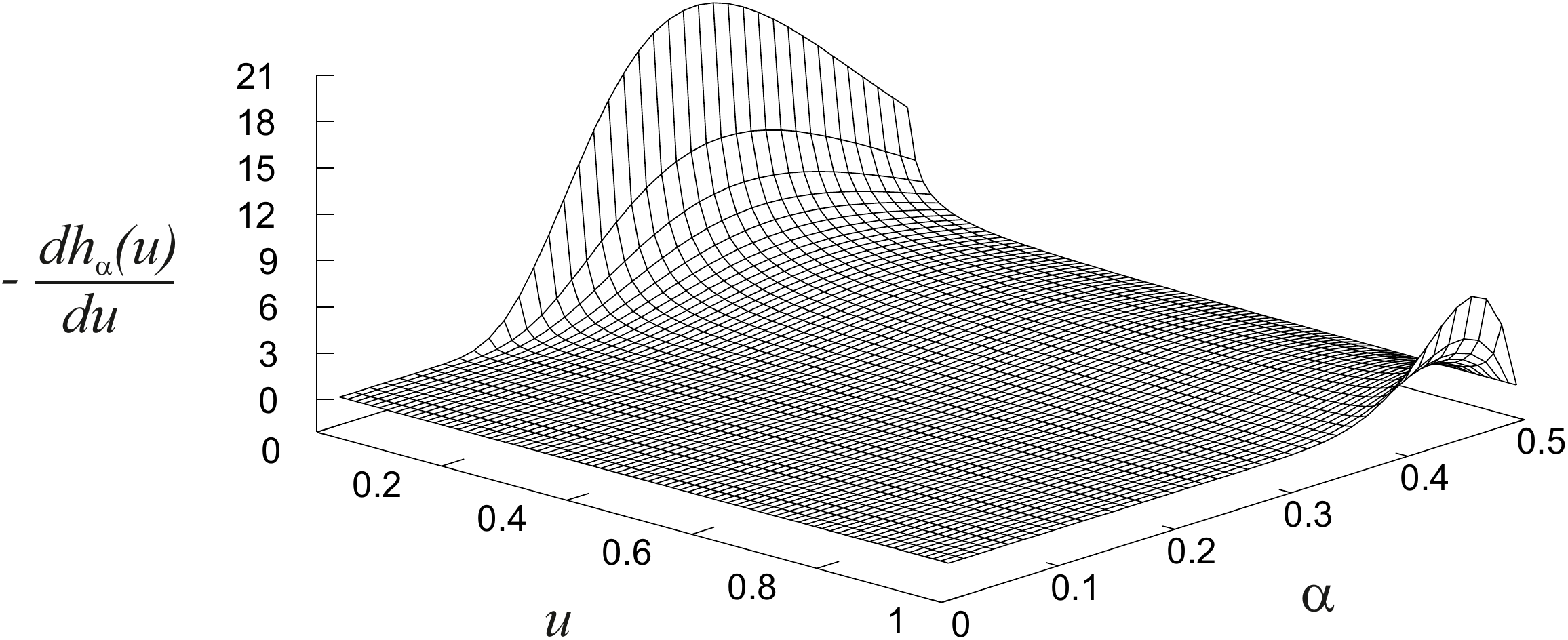}
\caption{Plot of $-\frac{d h_\alpha (u)}{d u}$ (cf. Eq. \eqref{eq:ineqosterok}), as a function of $u \in [0,1)$ and $\alpha \in (0,1/2]$, in the case of Jense-Shannon divergence. It can be seen that the quantity $h_\alpha(u)$ is nonincreasing for $\alpha \in (0,1/2]$. According to Theorem \ref{Ostertheorem} this result indicates that the quantity $[D_{JS}(P,Q)]^\alpha$ for $\alpha \in (0,1/2]$ is a metric.\label{fig:ineqosterok}}
\end{center}
\end{figure}

Figure \ref{fig:ineqosterok} shows a plot of the quantity $-dh_\alpha/du$, as a function of $u \in [0,1)$ and $\alpha \in (0,1/2]$ in the case of the JSD (cf. Eqs. \eqref{eq:metriccond} and \eqref{eq:ineqosterok}). It can be seen that the quantity $h_\alpha(u)$ is nonincreasing for $\alpha \in (0,1/2]$. Thus, according to Theorem \ref{Ostertheorem}, the quantity $[D_{JS}(P,Q)]^\alpha$ should be a metric for $\alpha \in (0,1/2]$. In what comes next, we state this hypothesis 
as Proposition \ref{prop:mainprop} and next we formally prove it.

\begin{proposition}
\label{prop:mainprop}
Let $D_{JS}(P,Q)$ denote the Jensen-Shannon divergence between two probability distributions $P = \{p_i \in \mathbb{R} \mid p_i \geq 0 ; \sum_{i=1}^n p_i = 1\}$ and $Q = \{ q_j \in \mathbb{R} \mid q_j \geq 0; \sum_{j=1}^n q_j = 1\}$. Then, the quantity

\begin{equation}
d_{\alpha}(P,Q) \doteq [D_{JS}(P,Q)]^\alpha
\end{equation}

\noindent is a (true) metric for all $\alpha \in (0,1/2]$. $\blacksquare$\\
\end{proposition}


\noindent\textbf{Proof:} The proof is based on theorem \ref{Ostertheorem} (cf. Sec. \ref{ciszarbasics}). Therefore, we need to prove that the function $h_{\alpha}(u)$ (cf. Eq. \eqref{eq:metriccond}) corresponding to the JSD is non-increasing for all $u\in [0,1)\subset \mathbb{R}$ and $\alpha \in (0,1/2]$. Thus, we shall analyze the sign of  the derivative of the function $h_{\alpha}(u)$. After some algebra, we obtain:

\begin{equation}
\frac{d h_\alpha(u)}{d u} = - \frac{(1-u^\alpha)^{\frac{1}{\alpha} - 1}.\left\{ u\log_2{u} +(u+u^\alpha)[1-\log_2(1+u)] \right\} }{2u[f_{JS}(u)]^2}\label{eq:ineqosterok}
\end{equation}

\noindent Clearly, the sign of $\frac{d h_\alpha}{d u}$ depends upon the sign of

\begin{equation}
 u\log_2{u} +(u+u^\alpha)[1-\log_2(1+u)]\label{sgnhu}
\end{equation}

\noindent because $(1-u^\alpha)^{\frac{1}{\alpha} - 1}$ and $u\,[f_{JS}(u)]^2$ are positive for  $u\in [0,1)$ and $\alpha>0$. Thus, in order to determine the conditions for $\frac{d h_\alpha}{d u}\leq 0$ we shall analyze the behavior of Eq. \eqref{sgnhu} as a function of $\alpha$ and $u$.

Starting from Eq. \eqref{sgnhu} we have the following sequence of inequalities:

\begin{align}
 u\log_2{u} +(u+u^\alpha)[1-\log_2(1+u)] \geq 0 \Leftrightarrow \\
 u^\alpha \left[ 1-\log_2(1+u) \right] \geq u\log_2(1+u) - u -u\log_2(u)  \Leftrightarrow  \\
 u^\alpha\log_2\left(\frac{2}{1+u}\right)  \geq u \log_2\left( \frac{1+u}{2u} \right) \Leftrightarrow \\
 u^\alpha  \geq u \frac{ \log_2\left( \frac{1+u}{2u}\right)}{\log_2\left(\frac{2}{1+u}\right)} = u \frac{ \ln\left( \frac{1+u}{2u}\right)}{\ln\left(\frac{2}{1+u}\right)}  \label{ineqhu}
\end{align}

In the case $\alpha = 1/2$ it has been proven that $[D_{JS}(P,Q)]^\alpha$ is a metric \cite{Oster1996,Endres2003}. Thus, inequality \ref{ineqhu} is satisfied for the particular case $\alpha = 1/2$.
Next, making use of the fact that for $u \in (0,1)$, if $\alpha \geq \beta$ then $u^\alpha \leq u^\beta$, we obtain the following inequality

\begin{equation}
u^\beta \geq u^{1/2} \geq u \frac{ \log_2\left( \frac{1+u}{2u}\right)}{\log_2\left(\frac{2}{1+u}\right)} ,
\end{equation}

\noindent for $\beta \in (0,1/2]$. As a consequence, $h_\alpha(u)$ turns out to be non-increasing in $[0,1)$ for all $\alpha \in (0,1/2]$. Therefore, 

\begin{align*}
d_\alpha(P,Q)=[D_{JS}(P,Q)]^\alpha
\end{align*}

\noindent is a metric for all $\alpha \in (0,1/2]$. $\square$

\begin{proposition}
For $\alpha \geq 1$, $[D_{JS}(P,Q)]^\alpha$ is not a metric. $\blacksquare$
\end{proposition}

\noindent\textbf{Proof:} For $\alpha \geq 1$ we need to prove that inequality \eqref{ineqhu} is violated. Making use of the fact that for $u \in (0,1)$, if $\alpha \geq 1$ then $u^\alpha \leq u$, $\ln(x)$ is a monotonically increasing function of $x$, and $(1+u)/2u > 2/(1+u) >1$ for $u \in (0,1)$, it follows that

\begin{equation}
 u^\alpha  < u\, \frac{ \ln\left( \frac{1+u}{2u}\right)}{\ln\left(\frac{2}{1+u}\right)},
\end{equation}

Thus, inequality \eqref{ineqhu} is not satisfied in this case. Therefore,  $[D_{JS}(P,Q)]^\alpha$ is not a metric for $\alpha \geq 1$. $\square$

This result is in agreement with the obtained by Khosravifard and co-workers in \cite{Khosra2007}. They conclude that Variational Distance (also known as Kolmogorov Distance) is the unique Csisz\'ar divergence which additionally is a metric \cite{Khosra2007}.

\begin{conjecture}
\label{metricconj}
For $1/2 < \alpha < 1$, $[D_{JS}(P,Q)]^\alpha$ is not a metric. $\blacksquare$
\end{conjecture}

In the case $\alpha \in (1/2 ,1)$ a number of counter--examples can be found showing that inequality \eqref{ineqhu} is violated for \textit{some} sub-intervals of values of $u \in (0,1)$, depending on the particular value of $\alpha \in (1/2 ,1)$. These results support the conjecture that $[D_{JS}(P,Q)]^\alpha$ is not a metric for $\alpha \in (1/2 ,1)$. However, so far we have not been able to prove this conjecture in a closed way.
In order to present some examples of the Conjecture \ref{metricconj}, we define the following function:

\begin{equation}
\Delta(u) \doteq u^\alpha - u\, \frac{ \ln\left( \frac{1+u}{2u}\right)}{\ln\left(\frac{2}{1+u}\right)}, \label{ineqhu2}
\end{equation}

\begin{figure}[h!]
\begin{center}
\includegraphics[clip,width=0.55\textwidth]{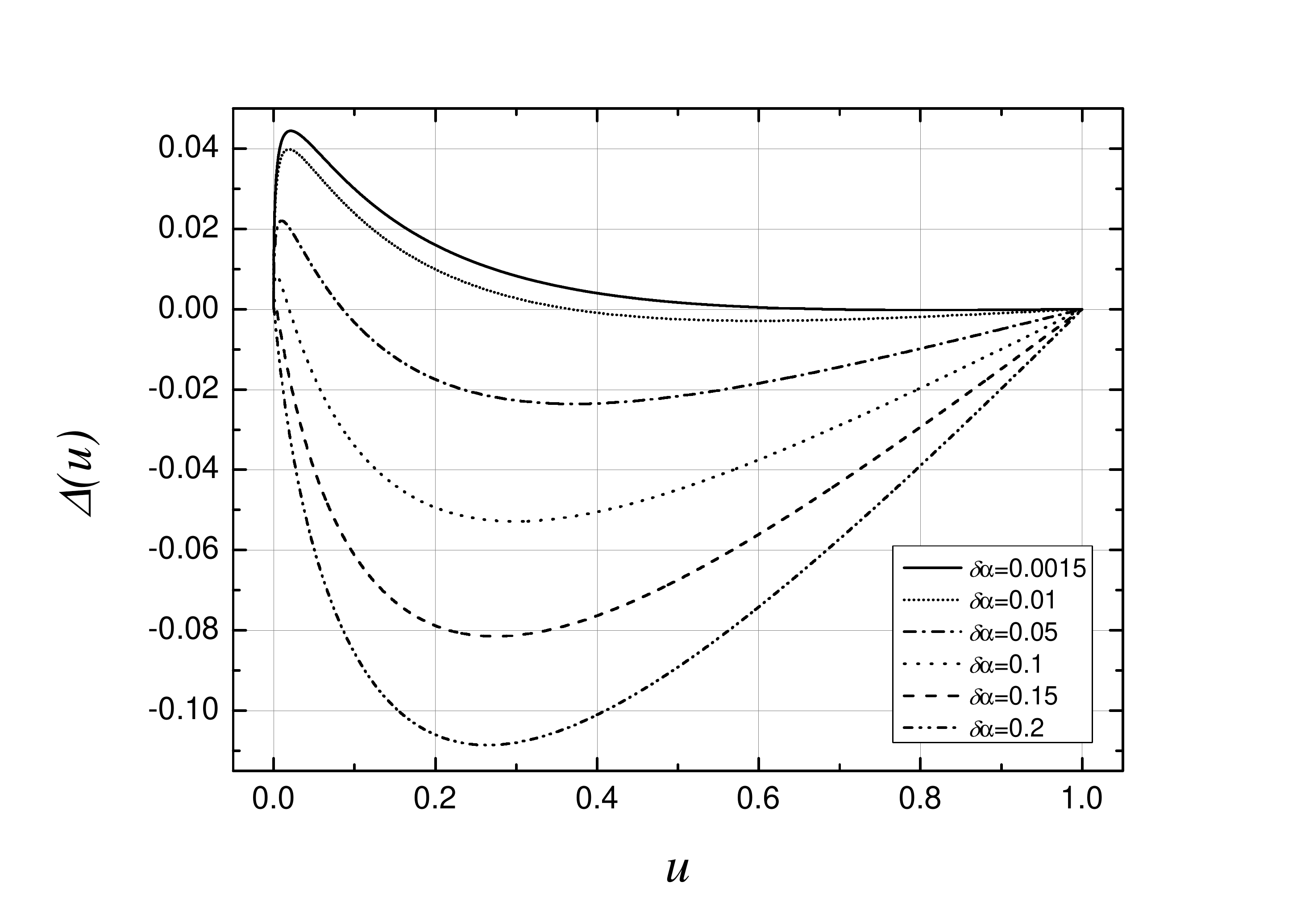}
\caption{Plot of $\Delta(u)$ (cf. Eq.\eqref{ineqhu2}) as a function of $u$ for some values of  $\alpha = 1/2+\delta\alpha$ with $\delta\alpha \in (0,1/2)$.  These results support the conjecture that
$[D_{JS}(P,Q)]^\alpha$ is not a metric for $\alpha \in (1/2,1)$.\label{fig:Deltau}}
\end{center}
\end{figure}

Figure \ref{fig:Deltau} shows some plots of the function $\Delta(u)$ for different values of the exponent $\alpha \doteq 1/2+\delta\alpha$ with $\delta \alpha \in (0,1/2)$. It can be seen that $\Delta(u) < 0$ for some subintervals of $u \in [0,1)$ showing that inequality \ref{ineqhu} can not be fulfilled for all values of $u\in [0,1)$. The size of the subinterval of values of $u$ for which $\Delta(u) < 0$ is enlarged as the value of $\delta\alpha$ is increased from $0^+$ to $1/2^-$. In fact, for values of $\delta \alpha \geq 0.2$ inequality \ref{ineqhu} begins to be unsatisfied for almost all values of $u\in [0,1)$. Thus, $[D_{JS}(P,Q)]^\alpha$ is not a metric for at least some values of the exponent $\alpha$ belonging to the interval $(1/2, 1)$.

\section{Applications\label{applications}}

In this section we briefly explore two possible applications of our main result, i.e., $[D_{JS}(P,Q)]^\alpha$ is a metric for $\alpha \in (0,1/2]$. In Sec. \ref{segmentation} we explore the segmentation of symbolic sequences whereas in Sec. \ref{Qrealm} we study the possibility to obtain a monoparametric family of metrics in the quantum realm.

\subsection{Segmentation of symbolic sequences\label{segmentation}}


A symbolic sequence is said statistically stationary if the frequency (probability) of occurrence of different symbols is the same along the entire sequence. In symbolic sequences resulting from real processes stationarity is not the most common situation. Thus, it is of great practical interest to provide a method to detect non-stationarity. In ref. \cite{Grosse2002} I. Grosse and co-workers developed a method based on the Jensen--Shannon divergence that allows the detection of changes in the statistical properties of symbolic sequences. This method and some modifications have been extensively used in the analysis of real and simulated time series \cite{Lamberti2003}. Typical applications range from the detection of epileptic crisis to the study of the alignment of the axis of an electric motor \cite{Mateos2017}. 

\begin{figure*}[htb]
\begin{minipage}[b]{.3\linewidth}
\begin{center}
 \subfigure{
 \includegraphics[scale =0.3] {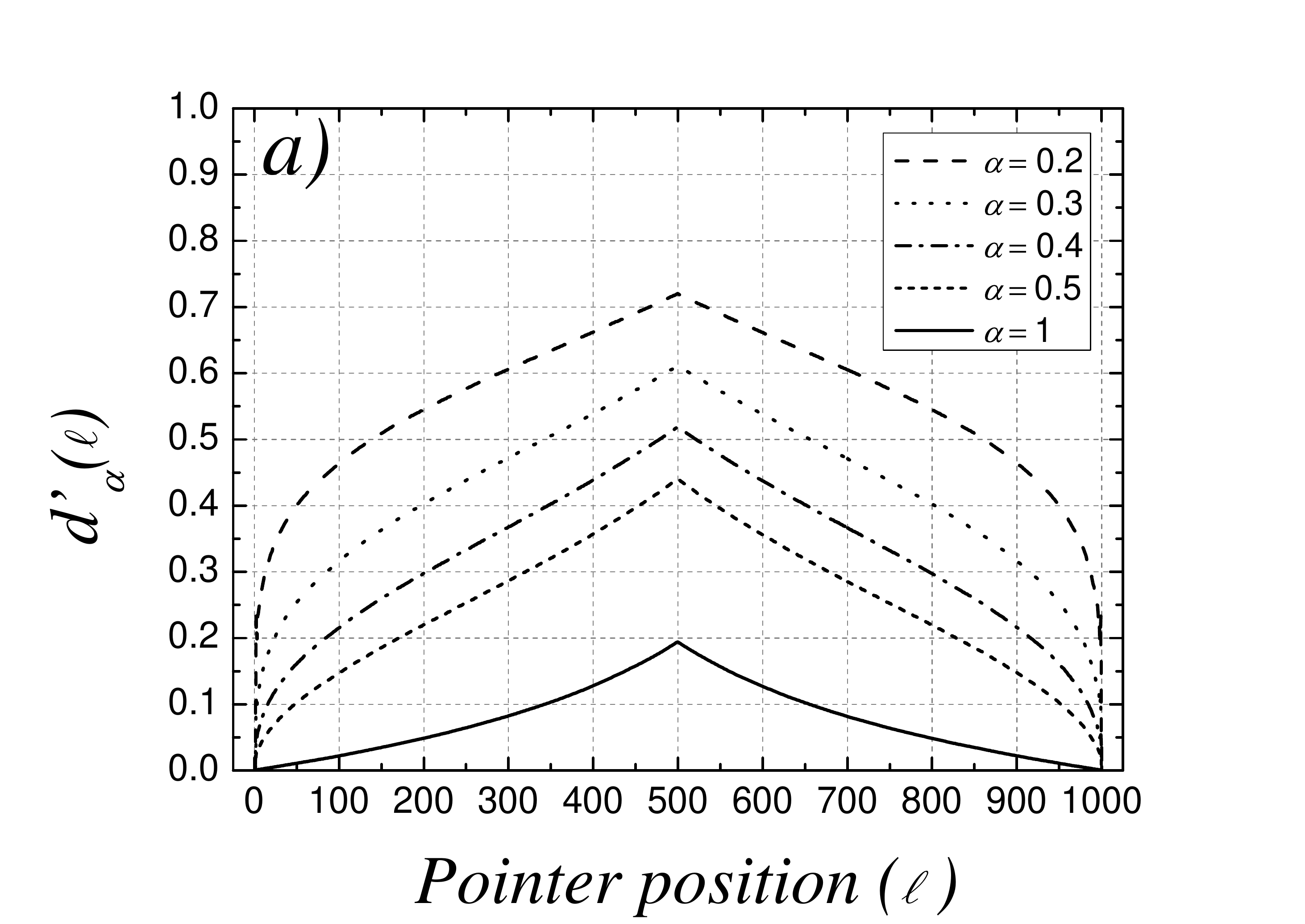}
   \label{fig:dalphax}
 }
\end{center}
\end{minipage}\hfill
\begin{minipage}[b]{.5\linewidth}
\begin{center}
\vspace{-0.2cm}
 \subfigure{
  \includegraphics[scale =0.3] {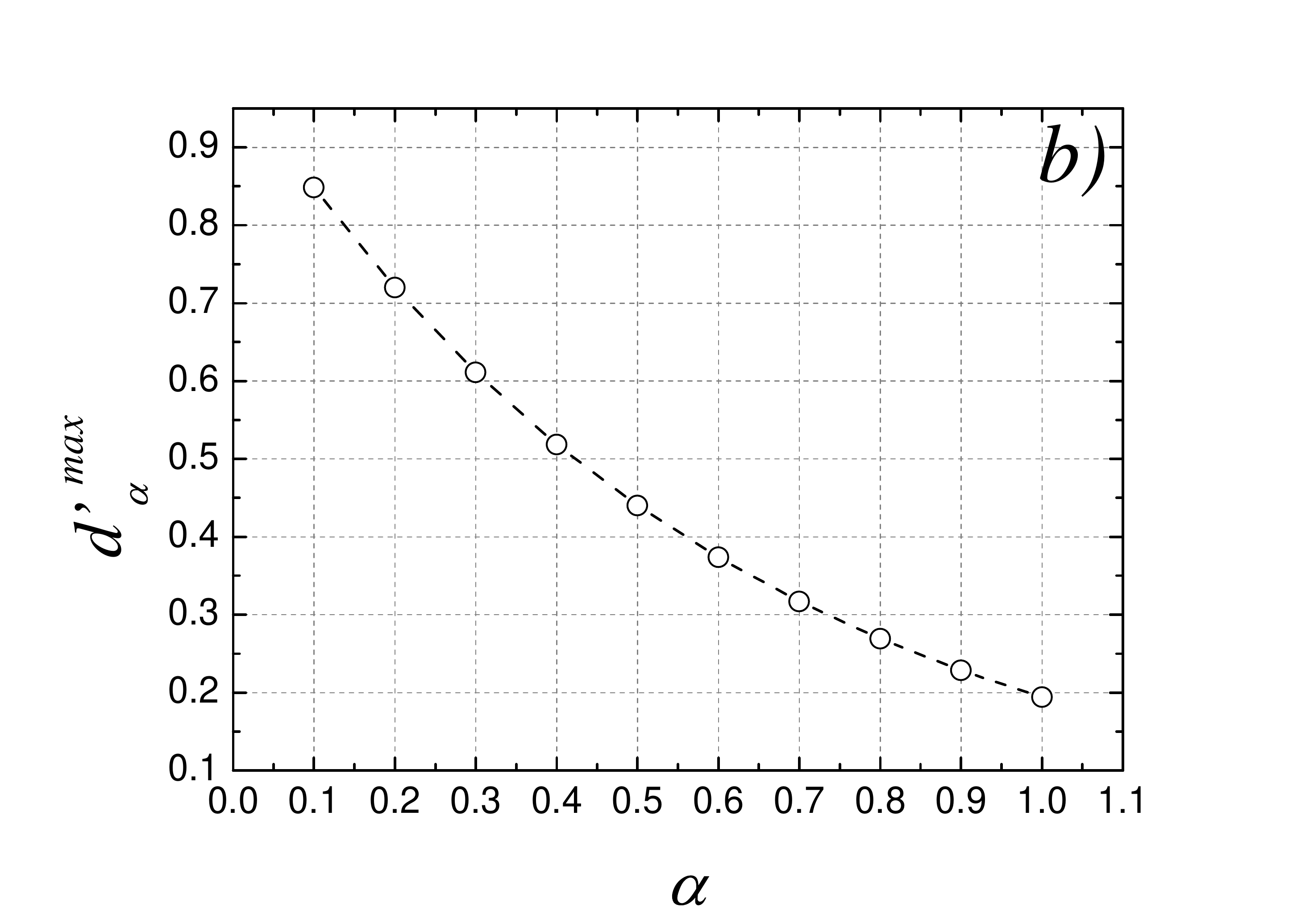}
   \label{fig:maxdalphax}
 }
\end{center}
\end{minipage}
\caption{a) Average results for the segmentation of $500$ binary sequences by means of $d_\alpha'$ (cf. Eq. \eqref{dprimealpha}). Each sequence of $1000$ symbols was generated by means of the Monte Carlo method. The maximum segmentation value occurs precisely for the cursor position $\ell=500$, i.e., the position where the probability distribution for the generated sequences changes from $\{p_0,p_1\}$ to $\{q_0,q_1\}$. b) Maximum segmentation value of $d_\alpha'$ as a function of the exponent $\alpha$. \label{fig:segmentresults}}
\end{figure*}

The monoparametric family of metrics introduced in the previous section (cf. Proposition \eqref{prop:mainprop}) motivated us to investigate the powers of $D^{(\pi_1,\pi_2)}_{JS}(P,Q)$ (cf. Eq. \eqref{JSDHVNgen}), in the range of the exponent $0 < \alpha \leq\frac{1}{2}$, as suitable quantities for studying the presence of non-stationarity in symbolic sequences. To this end, we implemented Monte Carlo simulations described as follows. We generated $500$ binary sequences of $1000$ symbols each one. The first $500$ symbols have a probability of occurrence $p_0=0.8$ and $p_1=0.2$ for symbols $0$ and $1$, respectively. The remaining $500$ symbols have a probability of occurrence $q_0=0.2$ and $q_1=0.8$. We introduced a mobile cursor along each sequence, denoting the position of this cursor with the letter $\ell$, $1\leq \ell \leq 1000$. For each position given by $\ell$ we define two weights $\pi_1= \frac{\ell}{1000}$ and $\pi_2 = \frac{1000-\ell}{1000}$. Then, we introduced the quantity

\begin{eqnarray}
d'_{\alpha}(\ell) &\equiv& \left[-\sum_{i=1}^2 (\pi_1 f_j + \pi_2 g_j) \log(\pi_1 f_j + \pi_2 g_j) \right.+\nonumber \\
&&\left.+\pi_1 \sum_{i=1}^2 f_j \log f_j + \pi_2 \sum_{j=1}^2 g_j \log g_j \right]^{\alpha}\label{dprimealpha}
\end{eqnarray}


\noindent where $f_j$ is the calculated frequency of occurrence of the symbol $j$ to the left of the cursor and $g_j$ is the calculated frequency of the symbol $j$ to the right of the cursor. Finally, we evaluated the average $d'_{\alpha}(\ell)$ over all the realizations of the sequences. In figure \ref{fig:segmentresults} we plot $d'_{\alpha}(\ell)$ as a function of $\ell$ for different values of $\alpha$. As the quantity $d'_{\alpha=1}(\ell)$ has been used in other works (see for example ref. \cite{Grosse2002}), for the sake of comparison, in figures \ref{fig:segmentresults} a) and b) we also show the results corresponding to $\alpha = 1$. In figure \ref{fig:segmentresults} a) we clearly observe that the maximum occurs at the position $\ell=500$, i.e., just the place where the probability distribution of the generated sequences changes from $\{p_0,p_1\}$ to $\{q_0,q_1\}$.\\
Now, when we face the problem of analyzing an unknown sequence, we do not have in general any \textit{a priori} knowledge about its possible stationary character, i.e., the location of probable segmentation points. Thus, when a likely segmentation point is found, the procedure needs to decide if such a point is \textit{statistically significant}, i.e., if it is significantly greater than expected by chance. Following ref. \cite{Grosse2002}, for an observed value $d'_{\alpha}=x$, the \textit{statistical significance} $s_{\alpha}(x)$ is defined as

\begin{equation}
s_{\alpha}(x) \doteq \mathrm{Prob}\{d'_{\alpha} \leq x\}\label{eq:defdprimesignif}
\end{equation}

\noindent Looking at eq. \eqref{eq:defdprimesignif}, $s_{\alpha}(x)$ is interpreted as the probability of obtaining $x$ or a lower value under the hypothesis that all subsequences are generated from the same probability distribution. As a result, following ref. \cite{Grosse2002}, the significance $s_{\alpha}(x)$ for large values of $N (\geq 10^2)$ can be estimated as \cite{Grosse2002}

\begin{equation}
s_{\alpha}(x) \sim \frac{\gamma(\nu/2, L\, (\ln 2) \, x^{1/\alpha})}{\Gamma(\nu/2)}\label{eq:dprimesignif}
\end{equation}

\noindent being $\nu = (N-1)(m -1)$, $L$ the length of the sequence, $m$ the number of subsequences and $N$ the number of symbols in the used alphabet, whereas $\gamma (a; x)$ and $\Gamma(a)$ represent the incomplete and complete gamma function, respectively \cite{Abram1972}.\\
Once we established the criterion to decide the statistical significance of potential segmentation points, the procedure for the analysis of stationary features in an unknown sequence proceeds as follows. We calculate the value of $d'_{\alpha}(\ell)$ for each position $\ell$ of the cursor along the entire sequence, as described earlier in this section. Next, we select the particular point at which $d'_{\alpha}$ reaches its maximum value $d'^{max}_{\alpha}$ and we compute its statistical significance $s^{max}_{\alpha}$. If the value $s^{max}_{\alpha}$ exceeds a given threshold $s_0$ the sequence is partitioned at this point and the procedure continues recursively for each one of the two resulting
subsequences. Otherwise, the sequence remains unpartitioned. The process is finished when none of the subsequent potential cutting points has a statistical significance exceeding $s_0$. In this case, we say
that the sequence has been segmented at a significance threshold $s_0$.\\
Figure \ref{fig:segmentresults} b) shows the dependence of the maximum value of $d'_{\alpha}$ found during the entire segmentation process of the simulated sequences described earlier in this section, as a function of $\alpha$. It can be seen that this maximum increases as the exponent $\alpha$ decreases. In addition, for a given threshold value $s_0$, it is straightforward to see from equation \eqref{eq:dprimesignif} that the statistical significance of $d'_{\alpha}$ monotonically increases as the value of $\alpha$ is decreased. This behavior is in complete agreement with the results showed in figure \ref{fig:segmentresults} b). Thus, 
the method based on the statistical significance of the maximum values of $d'_{\alpha}$ reached along the entire procedure turns out to be sensitive to the choice of the parameter $\alpha$. The results also suggest that a suitable election of the value of $\alpha$ might improve the proposed scheme. However, it is worth to mention that another methods can be found in literature for this kind of tasks. For example, a procedure based upon the calculation of the derivatives around the maximum value of a segmentation process is presented in refs. \cite{Nasci2014, Naran2017}. Further studies on this matter will be the subject of future research by the authors.

\subsection{Mono-parametric family of metrics in the quantum realm \label{Qrealm}}

It is well known that two quantum states can be discriminated unambiguously if, and only if, they are orthogonal. Thus, in the realm of Quantum Mechanics, distance measures need to be devised to allow us to determine how close two quantum states are from each other. At present, there is no general agreement about the use of a unique or ideal measure of distinguishability between quantum states. Moreover, different distance measures seem to be useful depending on the particular application, whether a theoretical one, like a bound of what can be physically feasible for a given process, or the output of a quantum protocol experimentally implemented. Therefore, from a conceptual point of view it is useful to develop new measures of distance between quantum states and then analyze their properties and possible applications.\\
On one hand, the probability distributions which determine the possible results to be obtained performing measurements on a quantum system with a state represented by a density matrix $\rho$ depend not only on $\rho$ but also on the set of measurements which can be performed on the system. On the other hand, the most general way of representing  measurements in the quantum realm is by means of the \textit{Positive Operator-Valued Measurement} (POVM) formalism \cite{Bengtsson2006,Helstrom1976,Nielsen2010}. Therefore, it is possible to extend the use of classical JSD into the quantum realm by means of two probability distributions defined as follows:

\begin{eqnarray}
P(\mathbb{E}, \rho) & = & \lbrace p_i \vert p_i = \Tr (E_i \rho)\label{eq:piprobs}\rbrace \\
Q(\mathbb{E}, \sigma) & = &  \lbrace q_i \vert q_i  = \Tr (E_i \sigma)\rbrace \label{eq:qiprobs}
\end{eqnarray}

\noindent where $\mathbb{E} = \{E_i\}_{i=1}^K$ represent some POVM mesurement ($\sum_{i=1}^M \mathbb{E}_i = \mathbb{I}$) \cite{Bengtsson2006,Nielsen2010}.

Thus, a distinguishability measure between the quantum states $\rho$ and $\sigma$ can be obtained from the classical JSD by assigning probabilities according to equations \eqref{eq:piprobs} and \eqref{eq:qiprobs}, and then optimizing over all possible POVMs. Since this procedure has the freedom of choosing the particular POVM distinguishing between the two probability distributions with more certainty, the following quantity can be introduced \cite{Majtey2005,Lamberti2008}:

\begin{equation}
D_{JS1}(\rho,\sigma)=\max_{\{\mathbb{E}_i \}} D_{JS}(p_i,q_i),
\label{eq:QJSD1}
\end{equation}

\noindent where the maximum is taken over the entire set of POVM's. Physically $D_{JS1}$ yields the best discrimination between the states $\rho$ and $\sigma$ that can achieved by means of measurements.
It is clear from the definition of $D_{JS1}(\rho,\sigma)$ that the results of our Proposition \ref{prop:mainprop} (cf. Sec. \ref{mainresult}) can be used to obtain a monoparametric family of metrics in the quantum realm in the form $[D_{JS1}(\rho,\sigma)]^\alpha$ with $\alpha \in (0,1/2]$.\\
Among the potential applications of the metric property of $[D_{JS1}(\rho,\sigma)]^\alpha$ we can mention its use as a tool for testing the convergence of iterative algorithms in quantum computation tasks \cite{Lamberti2008,Galindo2002,Osan2013} and to evaluate the performance of complex tasks of quantum information processing which can be decomposed into sequences of operations of lesser complexity \cite{Gilch2005}.

\section{Concluding remarks\label{concl}}

In this work we extended the previous and well-known result that the square root of the classical Jensen-Shannon divergence, i.e., $[D_{JS}(P,Q)]^{1/2}$, between two probability distributions $P$ and $Q$ is a metric by explicitly proving that it is possible to derive an entire monoparametric family of metrics from the classical JSD. Indeed, we demonstrated that the quantity $[D_{JS}(P,Q)]^\alpha$ is a metric for all $\alpha \in (0,1/2]$ in the sense that, in addition to the basic properties of a distance (Positive Definiteness, Symmetry and Identity of Indiscernibles, cf. Sec \ref{intrometrics}), it also satisfies the Triangle Inequality.  Furthermore, we explicitly demonstrated that the quantity  $[D_{JS}(P,Q)]^\alpha$ is not a metric for all $\alpha \geq 1$. The key aspects of our proofs were to consider the Jensen-Shannon divergence as a particular case of a Csisz\'ar divergence. We also conjectured that this quantity is not a metric for $\alpha \in (1/2,1)$ by providing some general examples supporting this hypothesis. However, so far we have not been able to obtain an analytical proof of this conjecture. This issue will be the subject of further studies. Also, motivated by our findings we briefly explored an application into the field of segmentation of symbolic sequences by introducing the $\alpha$--power of a quantity based on the generalized Jensen-Shannon divergence. Thus, the quantity $d'_{\alpha}$ (cf. sec. \ref{segmentation}) was studied as a tool for the detection of possible stationary features in symbolic sequences. The method used for segmentation is based on the statistical significance of the maximum values of $d'_{\alpha}$ reached along the entire procedure. On one hand, our results indicate that this method is sensitive to the particular choice of the parameter $\alpha$. On the other hand, they also suggest that a suitable election of the parameter $\alpha$ might improve this segmentation scheme. Additionally, we analyzed the possibility of extending the monoparametric family of metrics we derived from classical Jensen-Shannon divergence into the quantum realm. As a result, we also found a monoparametric family of metrics in the quantum realm, based upon the classical JSD, which can be derived from a quantum distance introduced in refs. \cite{Majtey2005,Lamberti2008} (cf. Eq. \eqref{eq:QJSD1}, Sec. \ref{Qrealm}). This last finding also deserves further study and the results will be presented elsewhere.\\

\section*{Acknowledgments}

The authors are grateful to F. Ost\"erreicher for gently providing a hardcopy of reference \cite{Kafka1991}. T.M.O and P.W.L are members of the National Research Council of Argentina (CONICET). D. G. B. has a fellowship from CONICET. The authors are grateful to Secretar\'{\i}a de Ciencia y T\'ecnica de la Universidad Nacional de C\'ordoba (SECyT-UNC, Argentina) and CONICET for financial support. 



\bibliography{JSmetricsbib}

\end{document}